\documentclass[11pt]{article}
\usepackage[a4paper,margin=1in]{geometry}
\usepackage[T1]{fontenc}
\usepackage{lmodern}
\usepackage{amssymb}
\usepackage{graphicx}
\usepackage{amsmath}
\usepackage{algorithm}
\usepackage{algpseudocode}
\usepackage[table]{xcolor}
\usepackage{multirow}
\usepackage{booktabs}
\usepackage{makecell}
\usepackage{tabularx}
\usepackage{adjustbox}
\usepackage{authblk}
\usepackage[colorlinks=true,linkcolor=blue,citecolor=blue,urlcolor=blue]{hyperref}
\title{SWIFT: A Handcrafted, Arithmetic-Minimal FPGA Engine for Real-Time Audio Deepfake Detection}
\author[1]{Chirantan Banerjee\thanks{Joint first authors.}}
\author[2]{Soumita Chatterjee\protect\footnotemark[1]}
\author[2]{Bhargab B. Bhattacharya}
\author[1]{Hafizur Rahaman}
\affil[1]{Department of Information Technology, Indian Institute of Engineering Science and Technology, Shibpur, Howrah, West Bengal 711103, India}
\affil[2]{School of VLSI Technology, Indian Institute of Engineering Science and Technology, Shibpur, Howrah, West Bengal 711103, India}
\affil[ ]{\texttt{\{chirantanbanerjee1208, csoumita45, bhargab.bhatta\}@gmail.com, rahaman\_h@yahoo.co.in}}
\date{}

\begin{document}
\maketitle

\begin{abstract}
Generative voice cloning threatens voice-based authentication, yet current deepfake detectors rely on power-hungry, high-precision models unsuited to edge hardware. We present SWIFT (Streaming Whiten-Integer Forensic Trust-Engine), a handcrafted detector that works in the high-frequency noise domain where synthesis artifacts concentrate. A first-order integer high-pass residual drives nine families of fixed-point descriptors, among them integer lifting-wavelet cepstral quantiles, Mahalanobis anomaly distances, Poincar\'{e} phase-space geometry, and phase jitter, fused by calibrated gradient-boosted trees and a sparse $L_1$-regularised meta-learner. Under leakage-controlled nested cross-validation, with one shared preprocessing chain for real and fake audio, this reference model reaches accuracy of up to 99.86\% and an Equal Error Rate (EER) as low as 0.03\% across eight benchmarks. Its integer front-end and a fixed-point linear classifier run on the programmable logic of a PYNQ-Z2 (Zynq-7020) board, with a 7.85\,$\mu$s per-frame datapath latency (127.4k frames/s) at 1.734\,W and 11.2\% LUT utilisation. Integer-only audio deepfake feature extraction is thus feasible at the edge without pretrained neural models or floating-point units.
\end{abstract}


\section{Introduction}

Audio deepfakes are especially dangerous because voice is tied to human trust, personal identity, and security authorization. A few seconds of reference audio now suffice to clone a voice, and attackers use such clones to bypass biometric banking checks, commit corporate fraud, and spread targeted disinformation. Since listeners can no longer reliably tell real from synthetic speech, automated detection has become a necessity for protecting digital communication.

Early audio deepfakes were built on Generative Adversarial Network (GAN)-based vocoders, benchmarked by WaveFake~\cite{frank2021wavefake} and LibriTTS~\cite{zen2019libritts}, with GAN-discriminator detectors~\cite{10.1145/3595353.3595883} and the ASVspoof 2021 challenge~\cite{10155166} serving as the main countermeasures of that era. As synthesis moved toward diffusion models and codec-language architectures such as VALL-E, these GAN-tuned detectors generalised poorly, pushing the field toward newer benchmarks such as MLAAD~\cite{10650962}, AUDETER~\cite{wang2025audeter}, and ShiftySpeech~\cite{garg2025shiftyspeech} built to capture these broader distribution shifts.

Detection methods have since diversified. Sundaram et al.~\cite{sundaram2024acoustic} paired handcrafted acoustic features with sequential models, and Doan et al.~\cite{doan2023btse} showed that breathing cues, which Text-to-Speech (TTS) generators rarely model, are a reliable liveness signal. Deep-learning detectors include depthwise-convolution transformers~\cite{pandey2025verivoice}, lightweight Convolutional Neural Networks (CNNs) on Mel-spectrograms~\cite{shafi2026deepfake}, dual self-supervised wavelet transformers~\cite{wang2025awaveformer}, and HuBERT-ECAPA embeddings~\cite{tahaoglu2025hubert}. More recent work covers codec-aware labelling~\cite{xiao2026how}, WavLM attentive fusion~\cite{10447923}, multi-stage frameworks~\cite{10888175}, codec-language countermeasures~\cite{10830534,10911062}, and multimodal audio-visual fusion~\cite{panja2025multimodal,11326621}.

Despite strong benchmark results, most of these methods depend on large neural encoders that cannot run on the embedded hardware of call centres, IoT devices, or mobile forensic kits. Hardware acceleration of deepfake detection has so far targeted quantised CNNs for image and video deepfakes on data-centre Field-Programmable Gate Array (FPGA) cards~\cite{bhilare2022deepfakecli}; to our knowledge, no handcrafted audio deepfake detector has been mapped to an embedded FPGA.

Deployed detectors are themselves machine-learning systems and inherit the threats that target such systems, including adversarial inputs~\cite{nafi2026dashmetaattackframeworksynthesizing}, Trojans in the machine-learning software and hardware stack~\cite{rahaman2026aispyparasitictrojansmachine}, and bit-flip faults in stored weights~\cite{rahaman2026sanaisecureeccdistributed}. Surveys~\cite{Rahaman2026} and runtime defences based on accelerator performance counters, including for audio inputs~\cite{rahaman2025runtimedetectionadversarialattacks,11471824}, address these threats, which makes compact, fully inspectable integer pipelines attractive for edge deployment.

We propose SWIFT, a neural-network-free detector that fuses nine Digital Signal Processing (DSP)-derived feature families through a calibrated stacking ensemble and runs its integer front-end on the programmable logic of a PYNQ-Z2 FPGA board. Our contributions are threefold. First, a nine-family, integer-friendly DSP descriptor of the acoustic noise floor, fused by Platt-calibrated LightGBM members and a sparse $L_1$ meta-learner, achieving up to 99.86\% accuracy and an Equal Error Rate (EER) as low as 0.03\% across eight benchmarks without any neural network. Second, a bit-exact fixed-point FPGA implementation of the front-end and a linear classifier on the PYNQ-Z2, with explicit hardware--software partitioning, a 7.85\,$\mu$s per-frame datapath latency (127.4k frames/s), 1.734\,W, and 11.22\% LUT utilisation---to the best of our knowledge, the first FPGA implementation of a handcrafted, neural-network-free audio deepfake detector. Third, a leakage-controlled evaluation with clip-level splits and nested cross-validation across TTS/codec, GAN-vocoder, ASVspoof, and mixed benchmarks, including a transform comparison and a comparison against five standard classifiers on identical features.

Section~\ref{sec:methodology} describes SWIFT and its hardware--software partitioning, Section~\ref{sec:results} reports the protocol, classifier comparison, cross-domain results, and the PYNQ-Z2~\cite{PYNQZ2} implementation, and Section~\ref{sec:conclusion} concludes.

\section{Methodology}
\label{sec:methodology}

\begin{figure}[t]
  \centering
  \includegraphics[width=0.54\linewidth]{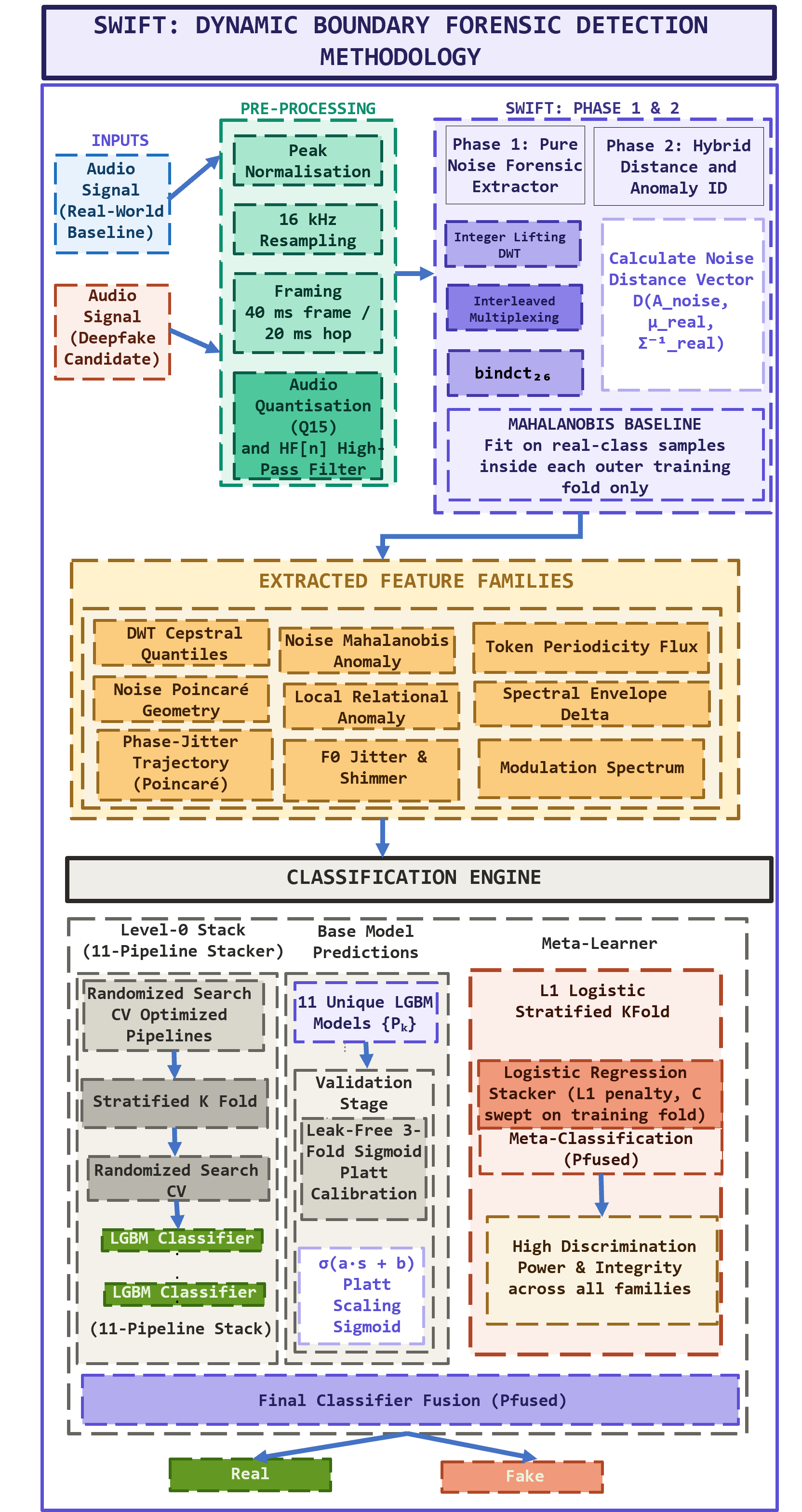}
  \caption{Block diagram of the proposed SWIFT pipeline.}
  \label{fig:block}
\end{figure}

SWIFT operates in the acoustic noise domain, where vocoders, diffusion synthesisers, GANs, and codec-language models leave faint structured traces that semantic features cannot expose. A noise residual isolates this domain, nine complementary DSP feature families capture distinct artifact classes, and a calibrated stacking ensemble fuses them (Fig.~\ref{fig:block}). The full ensemble is the software reference model behind all accuracy results; its integer front-end and a fixed-point linear classifier form the FPGA path of Section~\ref{subsec:fpga}.

\subsection{Preprocessing \& Signal Standardisation}
Real and deepfake audio are preprocessed identically. Each clip is converted to mono, peak-normalised to $[-1,1]$, and resampled to $f_s = 16$\,kHz through the shared chain of Section~\ref{subsec:protocol}, which keeps content below 7\,kHz. A sliding window of 40\,ms with a 20\,ms hop (50\% overlap) discretises the stream into frames, each quantised to Q15 fixed-point precision. To suppress semantic speech content, a first-order high-pass residual is extracted as
\begin{equation}
  hf[n] = x_{Q15}[2n{+}1] - x_{Q15}[2n],
  \label{eq:hf}
\end{equation}
since the difference of two Q15 samples spans $[-2,2)$. This is a first-order difference with magnitude response $2|\sin(\omega/2)|$ followed by $2\times$ decimation (the predict step of the integer Haar lifting scheme). It attenuates the low-frequency vocal-tract resonances that carry most phonetic content and exposes the high-frequency noise floor, where generative artifacts such as upsampling checkerboard patterns, phase-matching mismatches, and neural-vocoder rendering footprints accumulate. A first-order difference is chosen because it costs one integer subtraction per output sample and needs no multipliers, and Q15 arithmetic throughout makes software and hardware results bit-exact.

\begin{figure}[t]
  \centering
  \includegraphics[width=0.58\linewidth]{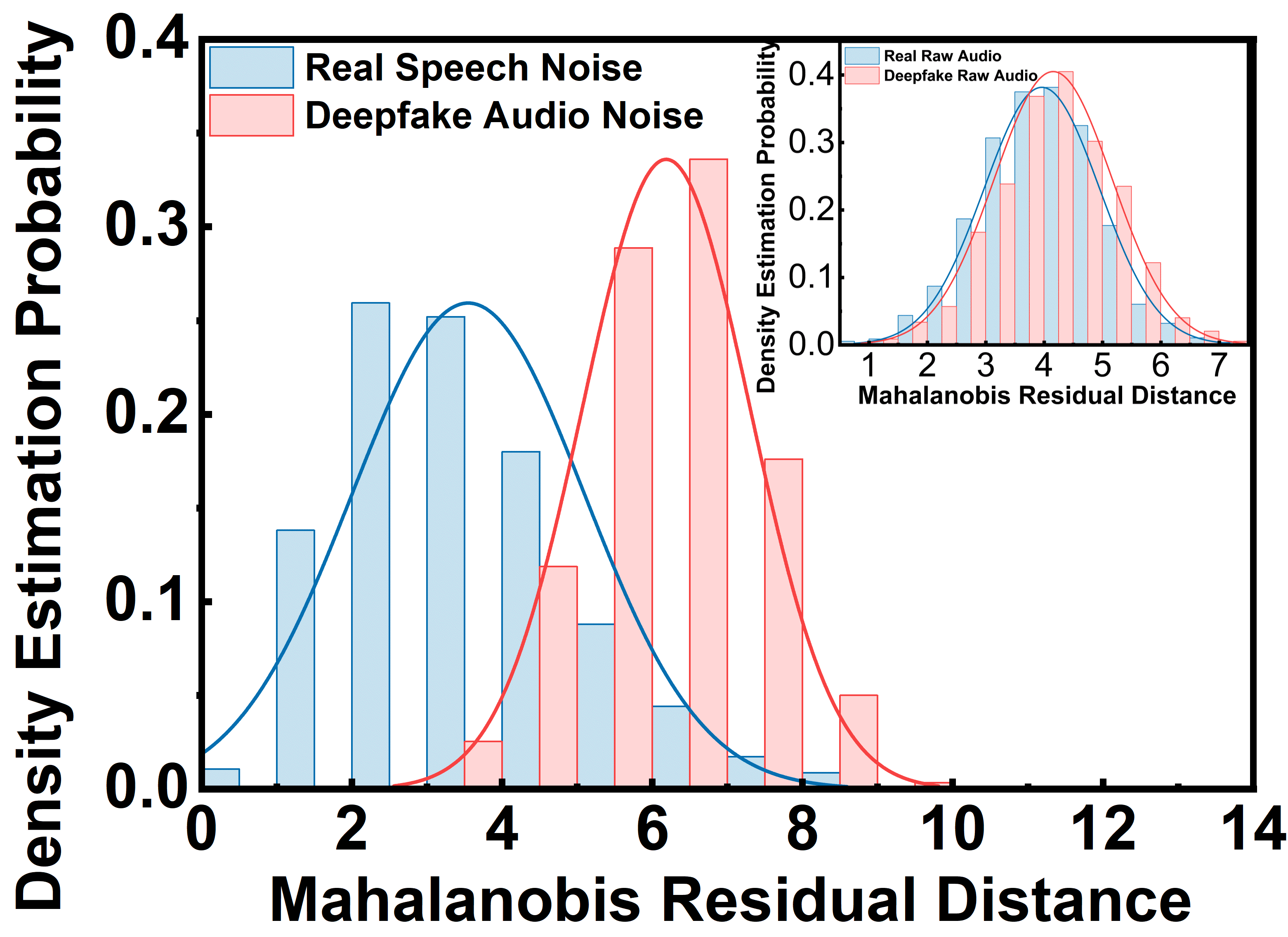}
  \caption{Isolated noise-domain Mahalanobis distances reveal distinct real vs.\ deepfake speech separation, unlike raw audio (inset).}
  \label{fig:residual_separation}
\end{figure}

Fig.~\ref{fig:residual_separation} validates this choice. The Mahalanobis distance~\cite{7727425} (Eq.~\eqref{eq:mahal}) measures how far a frame lies from the genuine-speech distribution while accounting for feature correlations. On the noise residual, real and deepfake speech form two clearly separated peaks that overlap only near a distance of 4, whereas on raw audio (inset) they overlap heavily, so the separation comes from isolating the residual rather than from the metric itself.

\subsection{Integer Lifting Wavelet and Global Anomaly Encoding}
\label{subsec:wavelet}
As shown in Fig.~\ref{fig:block}, feature extraction proceeds in two phases. In Phase~1, the noise residual of each frame is decomposed by an integer LeGall 5/3 lifting Discrete Wavelet Transform (DWT)~\cite{calderbank1998integer,1404263} in pure integer arithmetic. The first level's approximation is discarded, and its detail band $LH$ is split again into an approximation $HL$ and a detail $HH$. These detail bands are combined by asymmetric interleaved multiplexing into $\mathbf{c}=3\,\mathbf{s}_1+(\mathbf{s}_2\!\gg\!1)$, where $\mathbf{s}_1$ and $\mathbf{s}_2$ are the 128-sample, zero-padded sequences of every fourth coefficient of $LH$ and $HL$ ($\mathbf{s}_1$) and of $HH$ at offsets 0 and 2 ($\mathbf{s}_2$), and $\gg$ denotes an arithmetic right shift. The result is compressed by $\mathrm{BinDCT}_{26}$~\cite{844633} using only shifts and adds and summarised by four statistics (mean, median absolute deviation, 5th and 95th percentiles), which together form the 104-dimensional ($26\times4$) DWT cepstral descriptor, in the spirit of wavelet-based speech features~\cite{tufekci2000dwt}. Unlike irregular, physics-driven acoustic turbulence, block-based vocoder synthesis leaves periodic, structured spikes in the wavelet detail bands, artifacts that the integer DWT preserves and exposes. In Phase~2, a real-class baseline $(\boldsymbol{\mu}_\text{real}, \boldsymbol{\Sigma}_\text{real})$ fitted exclusively on authentic samples within each training fold scores every frame by its Mahalanobis distance,
\begin{equation}
  D_M(\mathbf{x}_t) = \sqrt{(\mathbf{x}_t-\boldsymbol{\mu}_\text{real})^{\!\top}
    \boldsymbol{\Sigma}_\text{real}^{-1}
    (\mathbf{x}_t-\boldsymbol{\mu}_\text{real})},
  \label{eq:mahal}
\end{equation}
where $\mathbf{x}_t$, $\boldsymbol{\mu}_\text{real}$, and $\boldsymbol{\Sigma}_\text{real}$ are the frame feature vector, real-class mean, and covariance, respectively. Ten summary statistics of $\{D_M(\mathbf{x}_t)\}$ capture the drift from the genuine microphone noise floor, a drift that current generators fail to avoid.

\subsection{The Remaining Feature Families}
\label{subsec:families}
The remaining seven families complete the 192-dimensional descriptor of Table~\ref{tab:families}. A local relational anomaly family scores every frame against the clip's own centroid, flagging recordings that are uniformly too smooth or that burst suddenly at a synthesis boundary. The noise residual is then embedded as a Poincar\'{e} plot~\cite{8965955} with short- and long-axis spreads
\begin{equation}
    SD_1 = \frac{\sigma(y[n+\tau] - y[n])}{\sqrt{2}}, \qquad
    SD_2 = \frac{\sigma(y[n+\tau] + y[n])}{\sqrt{2}},
    \label{eq:poincare}
\end{equation}
where $y[n]$ is the noise residual at sample $n$, $\tau$ is the embedding lag, and $\sigma(\cdot)$ is the standard deviation over $n$. $SD_1$ captures short-term sample-to-sample irregularity, while $SD_2$ tracks long-term attractor spread. Genuine phonation traces a rich, irregular attractor, whereas deterministic synthesisers produce attractors that are either too smooth or too erratic. A companion phase-jitter descriptor tracks frame-to-frame shifts in this geometry. The final four families address rhythm and voicing. Token periodicity flux exposes periodicities locked to the codec frame rate; $F_0$ jitter $J$ and shimmer $S$~\cite{farrus2007jitter} capture the vocal micro-tremor that synthesisers flatten or exaggerate; the log-spectral differences $\Delta S$ and $\Delta^2 S$ separate over-smooth transitions from impulsive boundaries; and the modulation-spectrum band ratio
\begin{equation}
    R_{\mathrm{mod}} = \frac{\sum_{f < 4} M(f)}{\sum_{4 \leq f < 20} M(f)},
    \label{eq:rmod}
\end{equation}
where $M(f)$ is the modulation-spectrum magnitude at modulation frequency $f$ (in Hz), contrasts slow sub-4\,Hz modulation energy with the 4--20\,Hz band that contains the natural 4--8\,Hz syllabic rhythm. The families are designed to be complementary, so that a fingerprint one family overlooks may be exposed by another.

\begin{table}[t]
\centering
\caption{Description of Feature Families}
\label{tab:families}
\setlength{\arrayrulewidth}{0.4pt}
\setlength{\tabcolsep}{4pt}
\small
\begin{tabular}{|c|l|c|l|}
\hline
\rowcolor[gray]{0.82}
\textbf{\#} & \textbf{Feature Family} & \textbf{Dim.} & \textbf{Target} \\
\hline
1 & DWT cepstral quantiles      & 104 & Wavelet boundary disc. \\
\hline
\rowcolor[gray]{0.95}
2 & Noise Mahalanobis anomaly   & 10  & Distribution shift \\
\hline
3 & Noise Poincar\'{e} geometry & 24  & Phase-space disruption \\
\hline
\rowcolor[gray]{0.95}
4 & Noise phase jitter          & 10  & Trajectory micro-var. \\
\hline
5 & Local relational anomaly    & 10  & Within-clip non-stat. \\
\hline
\rowcolor[gray]{0.95}
6 & Token periodicity flux      & 8   & Frame-rate rhythm \\
\hline
7 & $F_0$ jitter \& shimmer     & 8   & Prosodic micro-var. \\
\hline
\rowcolor[gray]{0.95}
8 & Spectral envelope delta     & 10  & Inter-frame change \\
\hline
9 & Modulation spectrum         & 8   & Syllabic rhythm \\
\hline
\rowcolor[gray]{0.82}
\multicolumn{2}{|l|}{\textbf{Total}} & \textbf{192} & \\
\hline
\end{tabular}
\end{table}

\subsection{Stacked Classification Engine}
\label{subsec:class}
The 192-dimensional fixed-point descriptor feeds the SWIFT stacked classification engine (Algorithm~\ref{alg:cws-he}), a two-level ensemble for the binary forensic decision. At Level-0, eleven parallel member pipelines are trained: one per feature family, one on the Principal Component Analysis (PCA)-whitened full concatenation, and one on the full concatenation without whitening. Each follows an identical four-step cascade of variance thresholding, standardisation, optional PCA decorrelation, and a class-balanced LightGBM classifier with extremely randomised splits, and each member is Platt-scaled via 3-fold cross-validation to yield calibrated probabilities. At Level-1, the eleven calibrated probability columns are assembled into a stacking matrix $\mathbf{P} \in \mathbb{R}^{N \times 11}$, where $N$ is the number of samples and each column holds one member pipeline's posterior probability, and fused by an $L_1$-regularised logistic regression,
\begin{equation}
\min_{\mathbf{w},\,b} \; \mathcal{L}_{\mathrm{CE}}(\mathbf{w}, b) + C^{-1}\|\mathbf{w}\|_{1},
\label{eq:l1_fuser}
\end{equation}
where $\mathbf{w} \in \mathbb{R}^{11}$ are the fusion weights, $b$ is the bias term, $\mathcal{L}_{\mathrm{CE}}$ is the cross-entropy loss, and $C^{-1}$ is the regularisation strength controlling sparsity. The $L_1$ penalty automatically suppresses redundant member pipelines, retaining only the most complementary probability streams for the final forensic decision. Every component is fitted inside the outer training folds of the protocol in Section~\ref{subsec:protocol}, and the Level-1 fuser is trained on out-of-fold Level-0 probabilities.

\begin{algorithm}[t]
\caption{SWIFT 2-Level Stacked Classification Engine.}
\label{alg:cws-he}
\begin{algorithmic}[1]
\Require feature matrix $\mathbf{F}\in\mathbb{R}^{N\times192}$, labels $y\in\{0,1\}^N$
\Ensure  stacking matrix $\mathbf{P}$; members $\mathcal{M}$; fuser $g$
\Statex \textbf{Level-0: Train 11 calibrated member pipelines}
\State $\mathcal{B}\gets$ 9 families $\cup\,\{\texttt{all},\texttt{all\_nopca}\}$
\ForAll{member $m \in \mathcal{B}$}
  \State prune features with variance $<10^{-5}$
  \State standardise: $\tilde{x}_j=(x_j-\hat\mu_j)/(\hat\sigma_j+\varepsilon)$
  \If{$m \neq \texttt{all\_nopca}$}
    \State PCA whiten, retain $99.5\%$ variance
  \EndIf
  \State fit LightGBM$_m$ (extra-trees, path-smooth, class-balanced)
  \State $\mathcal{M}[m]\gets\mathrm{PlattScale}(\text{LightGBM}_m,\,3\text{-fold})$
  \State $\mathbf{P}[:,m]\gets$ out-of-fold probabilities of $\mathcal{M}[m]$ on $\mathbf{F}$
\EndFor
\Statex \textbf{Level-1: Sparse logistic fusion}
\State fit $L_1$ fuser $g$ on $\mathbf{P}$: $\min_{\mathbf{w},b}\mathcal{L}_\mathrm{CE}+C^{-1}\|\mathbf{w}\|_1$
\State \Return $\mathbf{P},\,\mathcal{M},\,g$
\end{algorithmic}
\end{algorithm}

\subsection{Hardware--Software Partitioning and FPGA Deployment}
\label{subsec:fpga}
SWIFT is realised at two levels, summarised in Table~\ref{tab:partition}. The full nine-family stacking ensemble of Sections~\ref{subsec:wavelet}--\ref{subsec:class} is the \emph{software reference model}: it is trained offline and evaluated in software, and the accuracy figures in Tables~\ref{tab:dwt_justification}--\ref{tab:compare_prior} refer to it. The FPGA path, \emph{SWIFT-HW}, implements the integer front-end and a fixed-point linear classifier on the programmable logic (PL) of the PYNQ-Z2's Zynq XC7Z020. The LightGBM members, Platt calibration and $L_1$ fusion are \emph{not} mapped to the PL. A multiply--accumulate (MAC) datapath evaluates dot products, whereas decision-tree ensembles require comparator-based tree traversal, so SWIFT-HW instead uses a linear classifier trained offline on the 26 $\mathrm{BinDCT}_{26}$ coefficients, with int16 weights and an int32 bias. Mapping the tree ensemble to comparator logic is left for future work.

\begin{table}[t]
\centering
\caption{Hardware--Software Partitioning of SWIFT}
\label{tab:partition}
\setlength{\arrayrulewidth}{0.4pt}
\setlength{\tabcolsep}{3pt}
\footnotesize
\begin{tabularx}{\linewidth}{|>{\raggedright\arraybackslash}X|l|l|}
\hline
\rowcolor[gray]{0.82}
\textbf{Stage} & \textbf{Runs on} & \textbf{Arithmetic} \\
\hline
Noise residual (Eq.~\eqref{eq:hf}), integer lifting DWT, cross-multiplexing, $\mathrm{BinDCT}_{26}$ & PL & Integer add/shift \\
\hline
\rowcolor[gray]{0.95}
Linear classifier (SWIFT-HW) & PL & Fixed-point MAC \\
\hline
Weight loading, DMA, control, score read-back & PS (Cortex-A9) & -- \\
\hline
\rowcolor[gray]{0.95}
Families 2--9 & Software & Q15 fixed point \\
\hline
LightGBM, Platt, $L_1$ fusion; all training & Software & Floating point \\
\hline
\end{tabularx}
\end{table}

On the PL, Q15 frames arrive by Direct Memory Access (DMA) and stream through the residual of Eq.~\eqref{eq:hf} (one subtraction per sample pair), the two-level integer lifting DWT and cross-multiplexing (adders and shifters only), the strided $\mathrm{BinDCT}_{26}$ (shift--add butterflies), and the classifier MACs on the DSP slices, with no floating-point arithmetic. The ARM Cortex-A9 processing system (PS) loads the weights, issues DMA transfers and control handshakes, and reads back the score, driven from a Jupyter notebook with live audio. The IP core's outputs were compared with a bit-accurate software model of the datapath.

\section{Results and Discussion}
\label{sec:results}

\subsection{Experimental Protocol}
\label{subsec:protocol}
Each clip yields a single 192-dimensional descriptor, because every frame-level quantity (DWT coefficients, Mahalanobis distances, Poincar\'{e} spreads, jitter, and so on) is aggregated into clip-level statistics. All splits are therefore made at the clip level: the overlapping 40\,ms frames of a clip always fall on the same side of a split, so frame overlap cannot leak information from training to test data. Performance is estimated with nested cross-validation: an outer stratified 5-fold split for evaluation and, inside each outer training fold, a randomised hyper-parameter search with inner stratified cross-validation. Every fitted component (variance threshold, standardisation, PCA, the real-class Mahalanobis baseline, the LightGBM members, the 3-fold sigmoid Platt calibration, and the regularisation strength $C$ of the $L_1$ fuser) is fitted on the outer training fold only, and the Level-1 fuser is trained on out-of-fold Level-0 probabilities. Within each corpus the outer folds are stratified by label, so the same generators appear in training and test data; the results below therefore measure seen-generator detection, and generalisation to unseen generators is discussed as a limitation in Section~\ref{sec:conclusion}. Because SWIFT inspects the top of the spectrum, real and fake clips must share one decoding and resampling chain; otherwise the detector can learn storage traces instead of synthesis artifacts. The source corpora do differ in this respect: in MLAAD, whose fakes are 22.05\,kHz WAV files, the bona fide clips (LibriSpeech dev-clean) are 16\,kHz FLAC, and in CodecFake~\cite{wu2024codecfake} the genuine clips are 32-bit float while the fakes are 16-bit PCM at 16--44.1\,kHz. Every clip therefore passes through one chain: float-format audio is first rounded to 16 bits, all audio is resampled to 16\,kHz through a common 14\,kHz stage whose filters share the same pass ($<6$\,kHz) and stop ($>7$\,kHz) edges, and leading and trailing silence is trimmed. The sample counts in Table~\ref{tab:cross_domain} are clips; CodecFake combines all 44,242 genuine clips with 50,000 fakes. Reported values are computed on the pooled out-of-fold predictions.

\emph{Metric definitions.} The fake class is the positive class throughout. For each corpus, the fuser's out-of-fold probability of ``fake'' is pooled over all outer test folds into a single score vector, from which every metric is derived: AUC is the rank-based area under the ROC curve of this vector, EER is the operating point on the same ROC where the false-acceptance and false-rejection rates are equal, accuracy and recall (fake class) use the classifier's decision on the same samples, and F1 is the macro-average over the two classes (which is close to accuracy on balanced data). Because the ROC curve passes through $(\mathrm{EER},1-\mathrm{EER})$, every result must satisfy $\mathrm{AUC}\ge 1-\mathrm{EER}$; this check is enforced in the evaluation code, and all tables were regenerated from the same score vectors. The per-fold metric records are provided as supplementary material.

\subsection{Justification of Integer Lifting DWT on the Noise Domain}
\label{subsec:res_dwt}
\begin{table}[t]
\centering
\caption{Noise-Domain Transform Comparison}
\label{tab:dwt_justification}
\setlength{\arrayrulewidth}{0.4pt}
\setlength{\tabcolsep}{4pt}
\begin{tabular}{|l|c|c|c|}
\hline
\rowcolor[gray]{0.82}
\textbf{Transform (on Noise Residual)} & \shortstack[c]{\textbf{Accuracy} \\ \textbf{(\%)}} & \textbf{AUC} & \shortstack[c]{\textbf{EER} \\ \textbf{(\%)}} \\
\hline
Short-Time Fourier Transform & 97.50 & 0.9986 & 1.67 \\
\hline
\rowcolor[gray]{0.95}
Mel-Frequency Cepstral Coefficients & 75.83 & 0.8600 & 23.33 \\
\hline
Constant-Q Transform & 56.67 & 0.6594 & 40.00 \\
\hline
\rowcolor[gray]{0.95}
\textbf{Integer Lifting DWT (ours)} & \textbf{99.78} & \textbf{0.9995} & \textbf{0.05} \\
\hline
\end{tabular}
\end{table}
Table~\ref{tab:dwt_justification} compares four transforms applied identically to the noise residual under the same SWIFT pipeline. The Constant-Q Transform (CQT) performs worst (56.67\% accuracy, 40.00\% EER) because its logarithmically spaced bins smooth over the fine high-frequency spikes at vocoder boundaries, and Mel-Frequency Cepstral Coefficients (MFCC, 75.83\%) compress exactly the high-frequency content where artifacts accumulate. The Short-Time Fourier Transform (STFT) is competitive (97.50\%, 1.67\% EER), but its single fixed resolution cannot capture both fine token-boundary spikes and broader sub-band energy shifts from upsampling convolutions. The multi-resolution integer lifting DWT captures both, reaching 99.78\% accuracy and 0.05\% EER.

\subsection{Justification of the Classification Engine}
\label{subsec:res_feats}
Five well-known classifiers, a Support Vector Machine (SVM), a Multi-Layer Perceptron (MLP), Random Forest, XGBoost, and a single LightGBM, were trained on the same 192-dimensional features, outer folds and preprocessing as SWIFT (Table~\ref{tab:clf_compare}); these are the matched baselines of this paper. None exceeds 93\% accuracy or falls below 7\% EER: Random Forest is weakest (82.45\%, 17.49\% EER) and the MLP strongest (92.96\%, 7.18\% EER). Stacking the eleven calibrated pipelines lifts SWIFT to 99.78\% accuracy, an AUC of 0.9995, and 0.05\% EER. The jump from the single LightGBM (88.62\% accuracy) to the stack shows that no single model captures the complete picture. The contribution of each individual feature family is not isolated here and is left for future work.

\begin{table}[t]
\centering
\caption{Classifier Comparison on the ShiftySpeech Benchmark (same features and folds; macro-averaged F1)}
\label{tab:clf_compare}
\setlength{\arrayrulewidth}{0.4pt}
\setlength{\tabcolsep}{4pt}
\small
\begin{tabular}{|l|c|c|c|c|c|}
\hline
\rowcolor[gray]{0.82}
\textbf{Classifier} & \shortstack[c]{\textbf{Accuracy} \\ \textbf{(\%)}} & \shortstack[c]{\textbf{Recall} \\ \textbf{(\%)}} & \shortstack[c]{\textbf{F1-Score} \\ \textbf{(\%)}} & \textbf{AUC} & \shortstack[c]{\textbf{EER} \\ \textbf{(\%)}} \\
\hline
SVM (RBF) & 89.71 & 91.06 & 89.71 & 0.9685 & 10.42 \\
\hline
\rowcolor[gray]{0.95}
MLP (3-layer) & 92.96 & 93.48 & 92.96 & 0.9761 & 7.18 \\
\hline
Random Forest & 82.45 & 82.78 & 82.45 & 0.9070 & 17.49 \\
\hline
\rowcolor[gray]{0.95}
XGBoost & 88.51 & 88.98 & 88.51 & 0.9559 & 11.42 \\
\hline
LightGBM (single) & 88.62 & 89.00 & 88.62 & 0.9576 & 11.30 \\
\hline
\rowcolor[gray]{0.95}
\textbf{SWIFT (ours)} & \textbf{99.78} & \textbf{99.66} & \textbf{99.95} & \textbf{0.9995} & \textbf{0.05} \\
\hline
\end{tabular}
\end{table}

\subsection{Comparative Performance Analysis}
\label{subsec:res_main}
Table~\ref{tab:cross_domain} covers GAN-vocoder corpora (WaveFake~\cite{frank2021wavefake} and LibriTTS~\cite{zen2019libritts}), modern TTS and codec benchmarks (Audeter~\cite{wang2025audeter}, MLAAD~\cite{10650962}, CodecFake~\cite{wu2024codecfake}, ShiftySpeech~\cite{garg2025shiftyspeech}), ASVspoof~\cite{10155166}, and SONAR~\cite{li2410sonar}, which spans nine commercial and open-source TTS providers. On these four corpora SWIFT reaches 99.29--99.86\% accuracy with EERs of 0.05--0.66\%. LibriTTS yields a 0.03\% EER, while WaveFake is harder (90.42\%, 1.54\% EER) because of its diverse multi-vocoder composition; SONAR and ASVspoof reach 98.42\% and 97.84\%.

\begin{table}[t]
\centering
\caption{SWIFT Performance Across TTS/Codec, GAN-Vocoder, ASVspoof, and Mixed Benchmarks}
\label{tab:cross_domain}
\setlength{\arrayrulewidth}{0.4pt}
\setlength{\tabcolsep}{4pt}
\small
\begin{tabular}{|l|c|c|c|c|c|c|c|}
\hline
\rowcolor[gray]{0.82}
\textbf{Dataset} & \textbf{Type} & \textbf{Samples} & \shortstack[c]{\textbf{Accuracy} \\ \textbf{(\%)}} & \shortstack[c]{\textbf{Recall} \\ \textbf{(\%)}} & \shortstack[c]{\textbf{F1-Score} \\ \textbf{(\%)}} & \textbf{AUC} & \shortstack[c]{\textbf{EER} \\ \textbf{(\%)}} \\
\hline
Audeter & TTS/codec & 15,228 & 99.68 & 99.65 & 99.79 & 0.9896 & 0.29 \\
\hline
\rowcolor[gray]{0.95}
MLAAD & TTS/codec & 18,000 & 99.86 & 99.91 & 99.84 & 0.9898 & 0.12 \\
\hline
CodecFake & TTS/codec & 94,242 & 99.29 & 99.16 & 99.29 & 0.9889 & 0.66 \\
\hline
\rowcolor[gray]{0.95}
ShiftySpeech & TTS/codec & 281,945 & 99.78 & 99.66 & 99.95 & 0.9995 & 0.05 \\
\hline
WaveFake & GAN & 104,800 & 90.42 & 89.56 & 90.40 & 0.9235 & 1.54 \\
\hline
\rowcolor[gray]{0.95}
LibriTTS & GAN & 20,929 & 99.52 & 99.35 & 99.55 & 0.9992 & 0.03 \\
\hline
ASVspoof & TTS/VC & 121,461 & 97.84 & 97.12 & 97.48 & 0.9816 & 1.81 \\
\hline
\rowcolor[gray]{0.95}
SONAR & Mixed & 113,122 & 98.42 & 97.39 & 98.37 & 0.9741 & 1.27 \\
\hline
\end{tabular}
\end{table}

Against prior systems (Table~\ref{tab:compare_prior}), whose figures are quoted from the original papers and are therefore \emph{not} measured on identical data or folds, the Self-Supervised Learning (SSL) methods WavLM with a Multi-Fusion Attentive classifier (WavLM+MFA)~\cite{10447923}, XLS-R with contrastive learning (XLS-R+CL)~\cite{10888175}, and Wav2Vec~2.0 AASIST with Sharpness-Aware Minimisation (W2V2-AASIST+SAM)~\cite{10830534} plateau at 91--93\% accuracy despite large pretrained encoders, and a CNN on MFCCs~\cite{10911062} reaches 89.67\%, consistent with mel filterbanks discarding the high-frequency residual. The SWIFT reference model reaches 99.78\% accuracy with no pretrained encoder (the like-for-like comparison is Table~\ref{tab:clf_compare}), and its fixed-point path runs on FPGA hardware that cannot host any of these methods.

\begin{table}[t]
\centering
\caption{Comparison with Representative Prior Systems. Prior-system rows are quoted from the cited papers on the benchmark listed in the Benchmark column; the proposed row is the software reference model on ShiftySpeech}
\label{tab:compare_prior}
\setlength{\arrayrulewidth}{0.4pt}
\setlength{\tabcolsep}{3pt}
\small
\begin{tabular}{|l|c|c|c|c|c|c|c|}
\hline
\rowcolor[gray]{0.82}
\textbf{Methodology} & \textbf{Type} & \textbf{Benchmark} & \shortstack[c]{\textbf{Accuracy} \\ \textbf{(\%)}} & \shortstack[c]{\textbf{Recall} \\ \textbf{(\%)}} & \shortstack[c]{\textbf{F1-Score} \\ \textbf{(\%)}} & \textbf{AUC} & \shortstack[c]{\textbf{EER} \\ \textbf{(\%)}} \\
\hline
WavLM+MFA~\cite{10447923} & SSL & ASVspoof 2019 LA & 91.84 & 90.92 & 91.37 & 0.8470 & 0.52 \\
\hline
\rowcolor[gray]{0.95}
XLS-R+CL~\cite{10888175} & SSL & ASVspoof 2019 LA & 93.42 & 92.78 & 93.10 & 0.8517 & 0.46 \\
\hline
W2V2-AASIST+SAM~\cite{10830534} & SSL & CodecFake & 92.84 & 91.97 & 92.53 & 0.9416 & 0.62 \\
\hline
\rowcolor[gray]{0.95}
CNN+MFCC~\cite{10911062} & CNN & Fake-or-Real & 89.67 & 88.95 & 89.31 & 0.7597 & 0.31 \\
\hline
Proposed & DSP+Stack & ShiftySpeech & 99.78 & 99.66 & 99.95 & 0.9995 & 0.05 \\
\hline
\end{tabular}
\end{table}

\subsection{FPGA Implementation Results}
\label{subsec:res_fpga}
\begin{table}[t]
\centering
\caption{SWIFT-HW Implementation on the Zynq XC7Z020 at 50\,MHz}
\label{tab:resource}
\setlength{\arrayrulewidth}{0.4pt}
\setlength{\tabcolsep}{4pt}
\small
\begin{tabular}{|c|c|c|c|c|c|}
\hline
\rowcolor[gray]{0.82}
\textbf{Slice LUTs} & \textbf{Registers} & \textbf{DSPs} & \shortstack[c]{\textbf{PL latency} \\ \textbf{per frame}} & \shortstack[c]{\textbf{Throughput} \\ \textbf{(frames/s)}} & \textbf{Power} \\
\hline
\shortstack[c]{5,968 \\ (11.22\%)} & \shortstack[c]{8,647 \\ (8.13\%)} & \shortstack[c]{42 \\ (19.09\%)} & 7.85\,$\mu$s & 127.4k & 1.734\,W \\
\hline
\end{tabular}
\end{table}

The IP core occupies 11.22\% of the slice Look-Up Tables (LUTs) and 19.09\% of the DSP slices of the XC7Z020 (Table~\ref{tab:resource}). At 50\,MHz its datapath latency is 7.85\,$\mu$s per 40\,ms frame, i.e.\ 127.4k frames/s, over 2,500$\times$ the 50 frames/s that a 20\,ms hop requires. This figure covers the PL datapath only; DMA and PS overhead are not included. The board draws 1.734\,W.

\emph{Detection performance of the deployed classifier.} Table~\ref{tab:hw_acc} reports the SWIFT-HW linear classifier on the ShiftySpeech benchmark under the protocol of Section~\ref{subsec:protocol} (same outer folds, 26 $\mathrm{BinDCT}_{26}$ features of the PL front-end). The floating-point linear model is trained inside each training fold, and its weights are quantised to int16 (bias int32) and evaluated with the bit-accurate integer model of the datapath, which matched the IP-core outputs. The deployed classifier retains 98.45\% accuracy and 0.42\% EER, i.e.\ a loss of 1.33 accuracy points relative to the software ensemble, of which 0.35 points are due to quantisation alone. This trade-off is the price of moving from a 192-feature tree ensemble to a 26-feature multiplier-based classifier on the PL.

\begin{table}[t]
\centering
\caption{Detection Performance of the Deployed Classifier on ShiftySpeech (same folds)}
\label{tab:hw_acc}
\setlength{\arrayrulewidth}{0.4pt}
\setlength{\tabcolsep}{4pt}
\small
\begin{tabular}{|l|c|c|c|}
\hline
\rowcolor[gray]{0.82}
\textbf{Classifier} & \shortstack[c]{\textbf{Accuracy} \\ \textbf{(\%)}} & \textbf{AUC} & \shortstack[c]{\textbf{EER} \\ \textbf{(\%)}} \\
\hline
Linear, 26 BinDCT features, floating point & 98.80 & 0.9921 & 0.35 \\
\hline
\rowcolor[gray]{0.95}
\textbf{SWIFT-HW: linear, int16/int32 (deployed)} & \textbf{98.45} & \textbf{0.9904} & \textbf{0.42} \\
\hline
Software reference ensemble (192 features) & 99.78 & 0.9995 & 0.05 \\
\hline
\end{tabular}
\end{table}

\section{Conclusion}
\label{sec:conclusion}
This paper presented SWIFT, a handcrafted, neural-network-free audio deepfake detector that combines an integer lifting DWT cepstral transform with eight complementary descriptor families in a two-level calibrated stacking ensemble. The software reference reaches accuracy of up to 99.86\% (MLAAD) and an EER as low as 0.03\% (LibriTTS), with figures that compare favourably with the SSL and CNN results quoted in Table~\ref{tab:compare_prior} (reported on different benchmarks) without a pretrained encoder, and its fixed-point front-end and linear classifier run on the PYNQ-Z2 at 7.85\,$\mu$s per frame and 1.734\,W.

Four limitations remain: generators are shared between training and test folds, so the results measure seen-generator detection; robustness to compression, telephone channels, and noise (which alter the noise floor SWIFT relies on), to adversarial attacks, and cross-corpus transfer are untested; only the front-end and a linear classifier run on the PL, and the deployed classifier is evaluated on one benchmark only (Table~\ref{tab:hw_acc}) while the end-to-end (DMA and PS) latency is not reported; and the contribution of each feature family has not been isolated. Future work will add a leave-one-family-out ablation, evaluate the deployed fixed-point classifier on all benchmarks, map the tree ensemble onto comparator-based FPGA logic, evaluate unseen-generator, degraded-channel, and cross-lingual scenarios, and explore on-chip adaptation to emerging TTS architectures~\cite{xiao2026how}.

\section*{Acknowledgements}
This work has been financially sponsored by the \textbf{Anusandhan National Research Foundation (ANRF), Govt. of India} (Sanction Order No.\ \textbf{ANRF/PMP/2025/000071/ET}, dated 20 May 2026), and partially supported by the \textbf{Ministry of Electronics and Information Technology, Govt. of India}, under the \textbf{Chips to Startup (C2S) Programme}, vide Grant No.\ \textbf{EE-9/2/2021-R\&D-E}. We thank the \textbf{Indian Institute of Engineering Science and Technology, Shibpur}, Howrah, West Bengal, for supporting us with the necessary tools and systems to carry out the research work.

\bibliographystyle{unsrt}
\bibliography{ICISS}

\end{document}